\documentclass[twocolumn,aps,pra,superscriptaddress]{revtex4-1}
\usepackage{amsmath}
\usepackage{amssymb}
\usepackage{braket}
\usepackage{xcolor}
\usepackage{cancel}
\usepackage{mathtools}
\usepackage{mathbbol}
\usepackage{siunitx}

\newcommand{\omd}[0]{\omega_\mathrm{d} }
\newcommand{\ed}[0]{E_\mathrm{d}}
\newcommand{\sigrw}[0]{s_\phi}
\newcommand{\sigib}[0]{s_\omega}

\begin{document}

\title{Robustness of the Floquet-assisted superradiant phase and possible laser operation}
\author{Lukas Broers}
\affiliation{Center for Optical Quantum Technologies, University of Hamburg, Hamburg, Germany}
\affiliation{Institute for Quantum Physics, University of Hamburg, Hamburg, Germany}
\author{Ludwig Mathey}
\affiliation{Center for Optical Quantum Technologies, University of Hamburg, Hamburg, Germany}
\affiliation{Institute for Quantum Physics, University of Hamburg, Hamburg, Germany}
\affiliation{The Hamburg Center for Ultrafast Imaging, Hamburg, Germany}

\begin{abstract}
    We demonstrate the robustness of the recently established Floquet-assisted superradiant phase of the parametrically driven dissipative Dicke model, inspired by light-induced dynamics in graphene. 
    In particular, we show the robustness of this state against key imperfections and argue for the feasibility of utilizing it for laser operation.
    We consider the effect of a finite linewidth of the driving field, modelled via phase diffusion. 
    We find that the linewidth of the light field in the cavity narrows drastically across the FSP transition, reminiscent of a line narrowing at the laser transition.
    We then demonstrate that the FSP is robust against inhomogeneous broadening, while displaying a reduction of light intensity. 
    We show that the depleted population inversion of near-resonant Floquet states leads to hole burning in the inhomogeneously broadened Floquet spectra.
    Finally, we show that the FSP is robust against dissipation processes, with coefficients up to values that are experimentally available.
    We conclude that the FSP presents a robust mechanism that is capable of realistic laser operation.
\end{abstract}
\maketitle

\section{Introduction}

In the superradiant phase transition of the Dicke model \cite{Lieb73,Hioe73b} the ground state of a set of identical two-level systems (TLS) that are coupled to a cavity is accompanied by symmetry breaking and the emergence of a coherent photon state.
Realizations of the Dicke model, and consequently the superradiant phase transition, have been proposed \cite{Domokos02,Carmichael,Domokos10} and demonstrated experimentally \cite{Black03,Zhiqiang17,Esslinger10,Hemmerich1}.
The realization of the Dicke model in cavity-BEC setups leads to intricate non-equilibrium superradiant phases, which can appear in the presence of parametric driving of the coupling parameter \cite{Brandes1, Zilberberg, Ueda, Mathey1, Zhu19, Hemmerich4,Tuquero1,Cosme3,Cosme1,Hemmerich2,Cosme2}. %
Meanwhile, generalizations of the Dicke model have been studied to find rich phase diagrams that display superradiant phases, regular lasing and the unconventional counter-lasing \cite{Hioe73,Simons10,Keeling2,Ciuti14,Lesanovsky14,Zilberberg18,Carmichael18,Marino21,Parkins}.
These types of Dicke models are also referred to as driven, due to the tunability of the atom-photon processes \cite{IntroToDicke, Keeling3, Keeling1}.

\begin{figure}[ht!]
    \centering
    \includegraphics[width=0.904\linewidth]{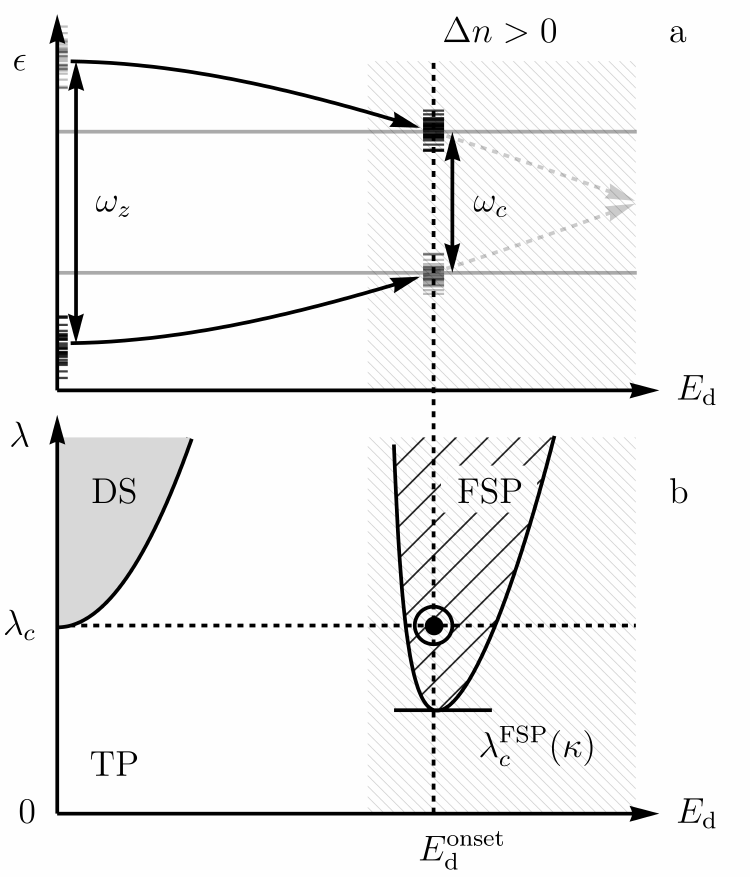}
    \caption{
    {\bf Mechanism of the Floquet-assisted superradiant phase.}
    (a) The bare energy levels $\omega_z$ of a collection of two-level systems are dressed via strong driving field strengths $\ed$ and deformed into Floquet states.
    Depending on the details of driving and dissipation, this process leads to population inversion in the Floquet states as indicated by the gray hatched areas.
    At Floquet energies that are resonant with the cavity frequency $\omega_c$, this population inversion is depleted and transferred into a coherent state in the cavity.
    (b) The phase diagram of the light field in the cavity shows the trivial phase (TP), the Dicke superradiant phase (DS) for small driving field strengths and the Floquet-assisted superradiant phase (FSP) for large driving field strengths around $\ed^\mathrm{onset}$.
    The coupling strength $\lambda_c^\mathrm{FSP}$ at which the FSP emerges depends on the cavity loss rate $\kappa$ that increases with $\kappa$.} 
    \label{fig0}
\end{figure}

The first studies on the superradiant phase were preceeded by studies on the closely related phenomenon of superradiance, introduced in the seminal work by Dicke \cite{Dicke}.
Unlike the superradiant phase, superradiance is a transient process of collective coherent spontaneous emission that can be engineered into continuous operation which results in superradiant lasing with ultranarrow linewidths \cite{Reynaud,Woerdman,Holland1,Thompson1,Chen,Hemmerich3,Thomsen,Holland2,Thompson2,Thompson3,Thompson4,Molmer1,Morigi1,Nicholson,Molmer4,Molmer3}.
The mechanisms behind regular lasing as well as superradiant lasing rely on incoherent driving which is realized via pumping into higher levels in order to create population inversion.
Superradiance has also been studied in the presence of coherent driving of the two-level systems \cite{Mollow1,Mollow77,Agarwal,Carmichael78,Walls80}.
This lead to studies on two-photon dressed-state lasers \cite{Lezama90,Lewenstein90,Berman91,Gauthier92,Zakrzewski911,Zakrzewski912,Zakrzewski913,Horak95,Gheri95} in particular in quantum dot system \cite{Shih07,Schoen1,Ilichev15,Shelykh1}, as a type of lasing without inversion \cite{Scully,Su92,Zhu92,LWIreview}.
Superradiance has also been studied in solid-state systems \cite{Kono16,Volz,Kay,Majer18}.
This relates to Floquet-engineering, which explores the possibilities of dynamically controlling properties such as band populations and topology \cite{Aoki09,Refael15,Esslinger17,Hsieh17,Budker,Thorwart1,Thorwart2,Petta1,FloquetReview,Yamanouchi1,MultimodePolariton,Lindner20}.

Generally speaking, one crucial feature of both regular and unconventional lasing is the generation of monochromatic scalable coherent emission that displays line narrowing below the intrinsic linewidth. 
Further, practical laser operation is expected to be stable in the presence of environmental factors.
Such factors include inhomogeneous broadening which in solid-state systems occurs due to material defects, while in gaseous setups it occurs due to the velocity distribution of the atoms.
This leads to spectral linewidths that exceed intrinsic linewidths due to shifted energy levels that would be degenerate in the absence of inhomogeneous broadening.

In recent work we have presented the Floquet-assisted superradiant phase (FSP) \cite{FSP} in a parametrically driven Dicke model in the presence of solid-like dissipation.
The underlying mechanism of the FSP consists of strong coherent driving of the TLSs which leads to Floquet states with energies that deviate from the bare transitions. 
These Floquet states can be tuned into resonance with a cavity while they simulatenously experience effective population inversion due to the interplay of coherent driving and dissipation.
This model is motivated by the demonstration of negative optical conductivities as a consequence of population inverted Floquet states in coherently driven graphene \cite{Broers}. 
Coupling the TLSs to the cavity results in the population inversion being depleted in order to sustain the oscillating coherent state in the cavity.
The mechanism of the FSP is illustrated in Fig.~\ref{fig0}. 
The regime in which inversion occurs as calculated in previous work \cite{FSP} is indicated with grey hatched lines.

In this paper, we demonstrate the robustness of the FSP.
We introduce a phase diffusion process and hence a finite linewidth in the driving field. 
We show that the transition into the FSP is accompanied by significant narrowing of the linewidth of the cavity light field such that the emergence of the FSP is stable in the presence of finite phase coherence of the driving field.
Further, we demonstrate that the FSP is stable in the presence of inhomogeneous broadening. 
Finally, we show that the FSP is robust with respect to the dissipation of the TLSs up to decay rates of the order of those in recent light-driven graphene experiments. 
We then identify the cavity loss rate as the most sensitive parameter, as the FSP vanishes comparatively rapidly as a function of cavity losses.
Overall, our results suggest the possibility of laser operation based on the FSP in a solid-state system, such as a two-band material, due to the form of the dissipative processes, the magnitude of the dissipation that we consider, the robustness against inhomogeneous broadening, and the strong line narrowing across the transition. 
This type of laser operation is distinct from regular lasing and superradiant lasing, but comparable to a modified type of dressed-state lasing in solids.

This work is structured as follows. 
In section II we describe the master equation of the parametrically driven dissipative Dicke model. 
In section III we introduce phase diffusion into the driving term of the TLS which leads to a broadened linewidth, that is overcome by the drastic line narrowing of the light field in the cavity.
In section IV we introduce inhomogeneous broadening into the TLSs which modifies the transition, as not all TLSs participate in the FSP. 
In section V we show the FSP transition as a function of the cavity loss rate as well as the TLS dissipation rates to show the robustness of the FSP.
In section VI we conclude our results and present an outlook for possible implementations of the FSP.

\section{Parametrically Driven Dicke Model}

We consider a dissipative Dicke model that is parametrically driven with circularly polarized light.
The Hamiltonian of this system is
\begin{eqnarray}
    \frac{1}{\hbar}H &= \sum_{j=1}^N \frac{\omega_z^j}{2}\sigma^j_z + \sigma_+^j  A_-(t)+ \sigma_-^j A_+(t)\nonumber\\
&   + \omega_c a^\dagger a+\frac{\lambda}{\sqrt{N}}\sum_{j=1}^N \sigma^j_x(a+a^\dagger)
    \label{Hmain}
\end{eqnarray}
with the driving field
\begin{equation}
    A_\pm(t) = \frac{\ed}{\omd } \exp\{\pm i\omd t\}.
    \label{vecpot}
\end{equation}
$\omd$ is the driving frequency and $\ed$ is the effective driving strength that takes the dimension of frequency squared. 
$\ed$ can be related to electric field strengths of driving terms in other systems that motivate the general form of this Hamiltonian. 
$\omega_c$ is the cavity frequency.
The frequencies $\omega_z^j$ are frequencies of the TLSs. 
We first choose these to be equal, i.e. $\omega_z^j = \omega_z$, and later consider a distribution of frequencies $\omega_z^j$, to model inhomogeneous broadening.
$\sigma_k^j$ is the $k$th Pauli-matrix acting on the $j$th TLS, where $\sigma_\pm^j=(\sigma_x^j\pm i\sigma_y^j)/2$.
$a^{(\dagger)}$ is the annihilation (creation) operator of the photon mode in the cavity.
$\lambda$ is the coupling strength between the TLSs and the cavity.
We consider a mean-field ansatz which separates the model into the two sub-Hamiltonians
\begin{align}
    \frac{1}{\hbar}H^j &= \frac{\omega_z^j}{2}\sigma_z^j + \sigma_+^j  A_-(t)+ \sigma_-^j A_+(t) +\frac{\lambda }{\sqrt{N}} \braket{a+a^\dagger}\sigma_x^j\\
    \frac{1}{\hbar}H_c &= \omega_c a^\dagger a +\frac{\lambda}{\sqrt{N}} \sum_{j=1}^N \braket{\sigma^j_x} (a+a^\dagger).\label{Hc}
\end{align}

Here $\braket{\sigma_x^j}$ and $\braket{a+a^\dagger}$ are the expectation values of the respective operators.
The dynamics generated by Eq.~\ref{Hc} are solved by a coherent state characterised by $\alpha=\braket{a}$ which acts as the order-paramater of superradiant phases. 
The dynamics of $\alpha$ are governed by the equation of motion 
\begin{equation}
    \dot\alpha = -(i\omega_c+\kappa)\alpha-i\frac{\lambda}{\sqrt{N}} \sum_{j=1}^N \braket{\sigma_x^j},
\end{equation}
where $\kappa$ is the cavity loss rate.
Additionally, we express the dynamics of the $j$th TLS via the Lindblad master equation which we write as
\begin{align}
    \dot \rho^j &= i[\rho^j, H^j] + \sum_{l} \gamma_l^j (L_l^j\rho^j L_l^{j, \dagger}-\frac{1}{2}\{L_l^{j, \dagger} L_l^j,\rho^j\}).
\end{align}
The Lindblad operators $L_l^j=\sigma_+^j,\sigma_-^j,\sigma_z^j$ are weighted by the dissipation coefficients $\gamma_l^j$ with $l\in\{+,-,z\}$ and act in the instantaneous eigenbasis of the $j$th TLS analogously to the method applied in previous works \cite{Nuske,FSP,Broers,Broers2}. 
The coefficients $\gamma_-^j$ and $\gamma_+^j$ describe spontaneous decay and excitation, respectively.
The coefficients $\gamma_z^j=\gamma_z$ describe dephasing.
This choice of dissipation has been shown to describe the dynamics in light-driven two-band solids \cite{Nuske}. 
Note that the temperature $T$ is encoded in the ratio of 
\begin{equation}
    \frac{\gamma_-^j - \gamma_+^j}{\gamma_-^j + \gamma_+^j} = \tanh\left(\frac{ \epsilon^j}{k_B T}\right),
\end{equation}
where the instantaneous eigenenergy scale $\epsilon^j$ of $H^j$ is roughly of the order of the $j$th TLS level spacing $\omega_z^j$.
Note that this suggests an ideal range of operation for the FSP. 
In terms of experimental feasibility it is desirable to keep $\ed$ small but $\frac{\epsilon^j}{k_B T}$ large.
This compromise is met up to room temperature for characteristic frequencies of the order of tens to hundreds of terahertz, which is in agreement with the motivational work on light-driven graphene \cite{McIver}. 
As an example throughout this work, we use $\omd=2\pi\times 48\si{\tera\hertz}$ and $\omega_z=2\pi\times 24\si{\tera\hertz}$ such that at room temperature $\tanh(\frac{\epsilon^j}{k_B T})\approx 1$. 
Therefore, we take $\gamma_+^j \approx 0$.

We further use $\omega_c=2\pi\times 12\si{\tera\hertz}$, $\gamma_-^j+\gamma_+^j=\gamma_-+\gamma_+=2\si{\tera\hertz}$, $\gamma_z=4\si{\tera\hertz}$, $\kappa=2\pi\times 120\si{\mega\hertz}$ and $\lambda_c=\sqrt{\omega_c\omega_z}/2\approx2\pi\times 8.5\si{\tera\hertz}$, where $\lambda_c$ is the critical coupling strength of the standard Dicke model.
The critical coupling $\lambda_c$ does not directly relate to the FSP, but it gives a readily comparable scale of the system parameters.
Unless stated otherwise, we take the driving field strength to be adjusted to the onset value of the FSP.
This means that the Floquet energies of the driven TLSs are resonant with the cavity frequency $\omega_c$, such that
\begin{equation}
    \ed=\frac{\omd^2}{2}\sqrt{\left(1-\frac{\omega_c}{\omd}\right)^2-\left(1-\frac{\omega_z}{\omd}\right)^2},
\end{equation}
as we have discussed in previous work \cite{FSP}.
In our example, with $\omega_c/\omd=1/4$ and $\omega_z/\omd=1/2$, this amounts to 
\begin{equation}
    \ed^\mathrm{onset} = \frac{\sqrt{5}}{8}\omd^2.
    \label{onset}
\end{equation}
The vertical line in Fig.~\ref{fig0} indicates the onset driving field strength.

\section{Phase Diffusion} 

\begin{figure}
    \centering
    \includegraphics[width=1.0\linewidth]{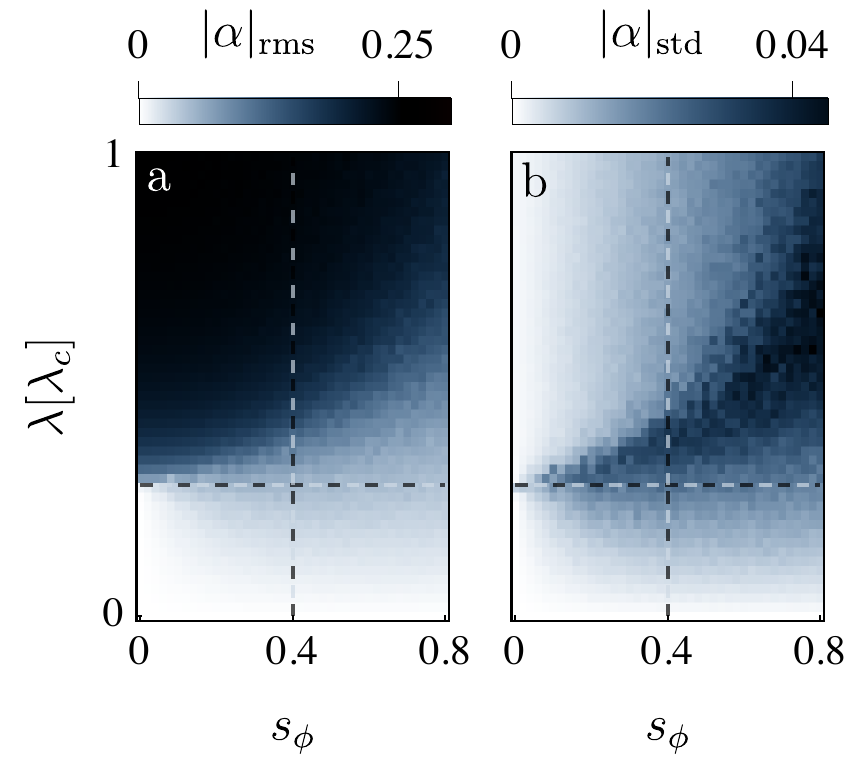}
    \includegraphics[width=1.0\linewidth]{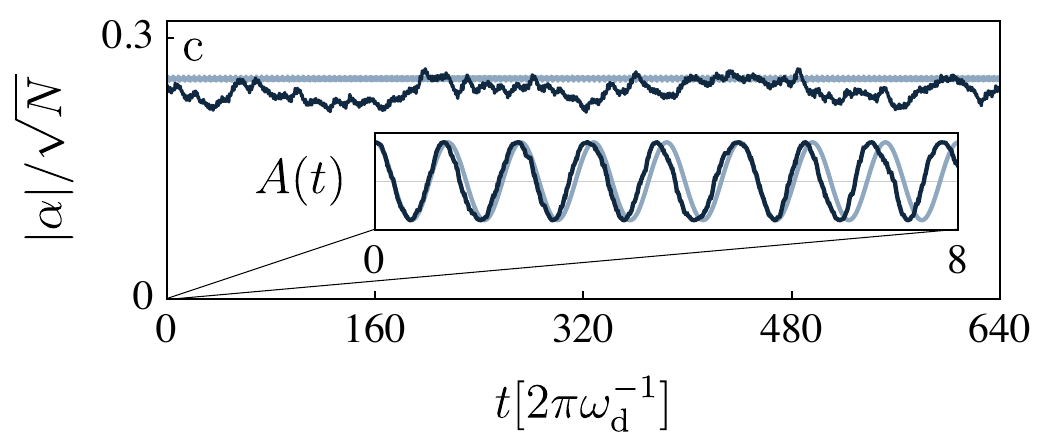}
    \caption{
    {\bf Light field fluctuations across the FSP transition.}
    Panels (a) and (b) show the root-mean-square and the standard deviation of the light field amplitude across the FSP transition as a function of the phase diffusion standard deviation $\sigrw$.
    The horizontal line indicates the coupling strength at which the FSP transition occurs in the absence of phase diffusion, i.e. $\sigrw=0$. 
    Panel (c) shows the amplitude of a single-trajectory of the light field for $\sigrw=0$ (blue) and $\sigrw=0.4$ (black),
    as well as an inset of the driving field for the same values of $\sigrw$.}
    \label{diffusionscan}
\end{figure}

As a first metric for robustness, we consider the influence of a finite linewidth of the driving field on the linewidth of the cavity light field in the FSP.
For this purpose we introduce a Gaussian random walk $\phi(t)$ that models phase diffusion in the driving field
\footnote{We write the discretized random walk as
\begin{align}
    \phi_n = \sum_{i=0}^n \mathcal{N}\left(0, \frac{\sigrw^2}{n_T} \right),
\end{align}
where $\mathcal{N}(\mu,\nu)$ indicates a randomly sampled value from the normal distribution with mean $\mu$ and variance $\nu$.
$n_T$ is the amount of time-steps per driving period, such that $\phi(t)$ is determined by interpolation of $\phi_n$.}, rather than the monochromatic driving field described in Eq.~\ref{vecpot}. 
The standard deviation of $\phi(t)$ after one driving period $2\pi\omd^{-1}$ is given by $\sigrw$.
This corresponds to a linewidth of $\Delta\omega = \omd \frac{\sigrw}{2\pi}$ in the driving field which we now write as
\begin{equation}
    A_\pm(t) = \frac{\ed}{\omd}\exp\{\pm i(\omd t + \phi(t))\}.
\end{equation}

In Fig.~\ref{diffusionscan}~(a), we show the root-mean-square of the light field amplitude $|\alpha|_\mathrm{rms}$ across the FSP transition at the onset value of $\ed=\ed^\mathrm{onset}$, see Eq.~\ref{onset}, as a function of $\lambda$ and $\sigrw$. 
We see that the FSP transition is stable against the fluctuations of the driving field, i.e. as a function of $\sigrw$, however the FSP regime is shifted towards larger values of $\lambda$, and the transition regime displays stronger fluctuations.
Since the phase diffusion broadens the linewidth of the driving field, the TLSs develop amplitude at many frequencies including the cavity frequency $\omega_c$ which leads to residual occupations in the cavity that contribute to the broadened FSP transition. 

In Fig.~\ref{diffusionscan}~(b), we show the standard deviation $|\alpha|_\mathrm{std}$ of the light field amplitude as a function of $\lambda$ and $\sigrw$.
Close to the FSP transition, the amplitude that we show in Fig.~\ref{diffusionscan}~(a) displays large fluctuations, that are suppressed both in the FSP and the trivial phase.
With increasing phase diffusion, the steady state in the FSP shows an increasing standard deviation and the sharp feature in $|\alpha|_\mathrm{std}$ across the transition broadens.
For intermediate values of $\sigrw$, the increased standard deviation indicates the shifted location of the FSP transition.

In Fig.~\ref{diffusionscan}~(c), we show the amplitude of the light field in the FSP at $\lambda=\lambda_c$ for $\sigrw=0$ and $\sigrw=0.4$ on long time scales. 
The case of $\sigrw=0.4$ corresponds to a significantly broadened driving field and hence the amplitude of the light field in the FSP jitters considerably.
This case corresponds to the vertical dashed lines in Figs.~\ref{diffusionscan}~(a) and (b).
The inset shows the effect of the phase diffusion on the driving field.

\begin{figure}
    \centering 
    \includegraphics[width=1.0\linewidth]{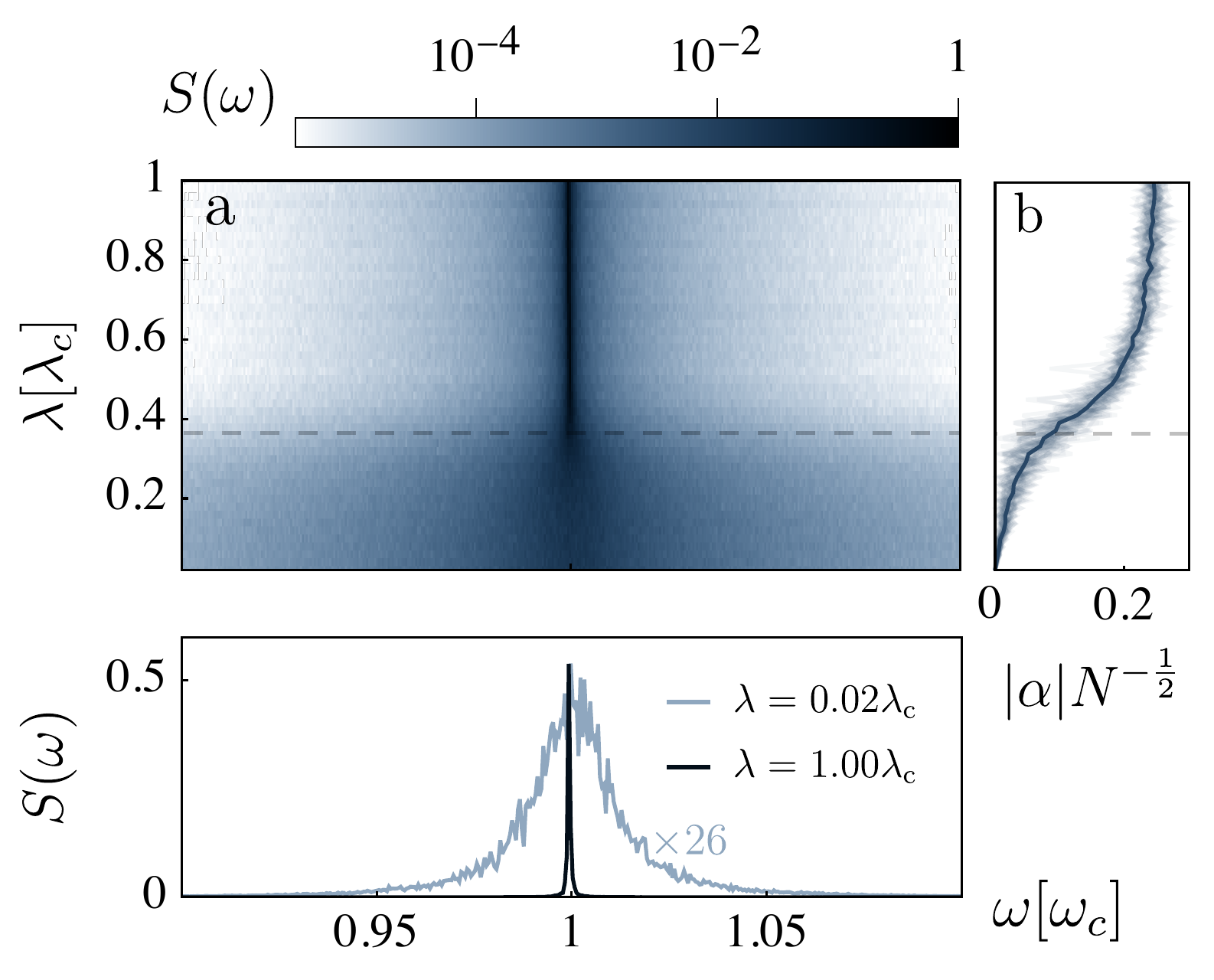}
    \caption{
    {\bf Linewidth narrowing in the FSP.}
    Panel (a) shows the power spectrum averaged over 50 phase diffusion trajectories as a function of the coupling strength $\lambda$ for $\sigrw=0.4$ on a logarithmic scale.
    Across the FSP transition indicated by the dashed line, the linewidth of the light field in the cavity narrows drastically.
    Panel (b) shows the light field amplitude across the FSP transition for the individual trajectories in light colors and their mean in solid dark blue.
    Panel (c) shows the power spectra for $\lambda = 0.02\lambda_c$, rescaled by a factor of $26$ for comparison, and $\lambda=\lambda_c$. 
    }
    \label{narrowing}
\end{figure}

Next we present the line narrowing of the light field across the FSP transition in the presence of phase diffusion.
For this purpose we consider the power spectrum 
\begin{equation}
    S(\omega) = \frac{|\hat\alpha(\omega)|^2}{\int_\mathbb{R} |\hat\alpha(\omega)|^2 d\omega},
\end{equation}
where $\hat\alpha(\omega)$ is the Fourier transform of $\alpha(t)$. 

In Fig.~\ref{narrowing}~(a) we show $S(\omega)$ at the onset driving field strength with $\sigrw=0.4$ and averaged over 50 phase diffusion trajectories. 
Prior to the FSP transition that occurs at approximately $\lambda=0.37\lambda_c$, there is a broadened signal at the cavity frequency with a linewidth that is given by the cavity loss rate $\kappa$.
Across the transition, the linewidth narrows drastically as the occupation of the cavity mode increases. 
We show the corresponding light field amplitude $|\alpha|$ in Fig.~\ref{narrowing}~(b) as a transposed plot for the same set of sampled phase diffusion trajectories in light colors, and their mean as a dark solid line.

In Fig.~\ref{narrowing}~(c) we show the power spectrum $S(\omega)$ prior to ($\lambda=0.02\lambda_c$) and past ($\lambda=\lambda_c$) the FSP transition, rescaled for comparison.
The line narrowing is clearly visible as the full-width half-maximum reduces drastically.
The narrowing of the linewidth past the transition corresponds to the coherence times increasing beyond those of the driving field and those given intrinsically by the cavity. 
This is a hallmark of lasing states.

\section{Inhomogeneous Broadening}

As a second metric for robustness we consider the consequences of inhomogeneous broadening on the FSP transition.
In order to include inhomogeneous broadening, we consider a distribution of detuned two-level spacings $\omega_z^j$ that have the average frequency $\omega_z$.
In this model the index of the TLSs is arbitrary, so we label them by their energy for convenience.
The relevant quantity for broadening effects is the mode density of the TLSs.
We consider $N=100$ TLSs with normal distributed energy detuning around $\omega_z$ with a relative standard deviation of $\sigib$.
For this purpose, we set the $j$th energy level to 
\begin{equation}
    \omega_z^j = \omega_z \left(1 + \sigib\mathrm{erf}^{-1}\left[\frac{2j}{N+1}-1\right]\right),
    \label{erf}
\end{equation}
where $\mathrm{erf}^{-1}$ is the inverse error function, such that the energy levels are evenly distributed as desired without random sampling.

\begin{figure}
    \centering
    \includegraphics[width=1.0\linewidth]{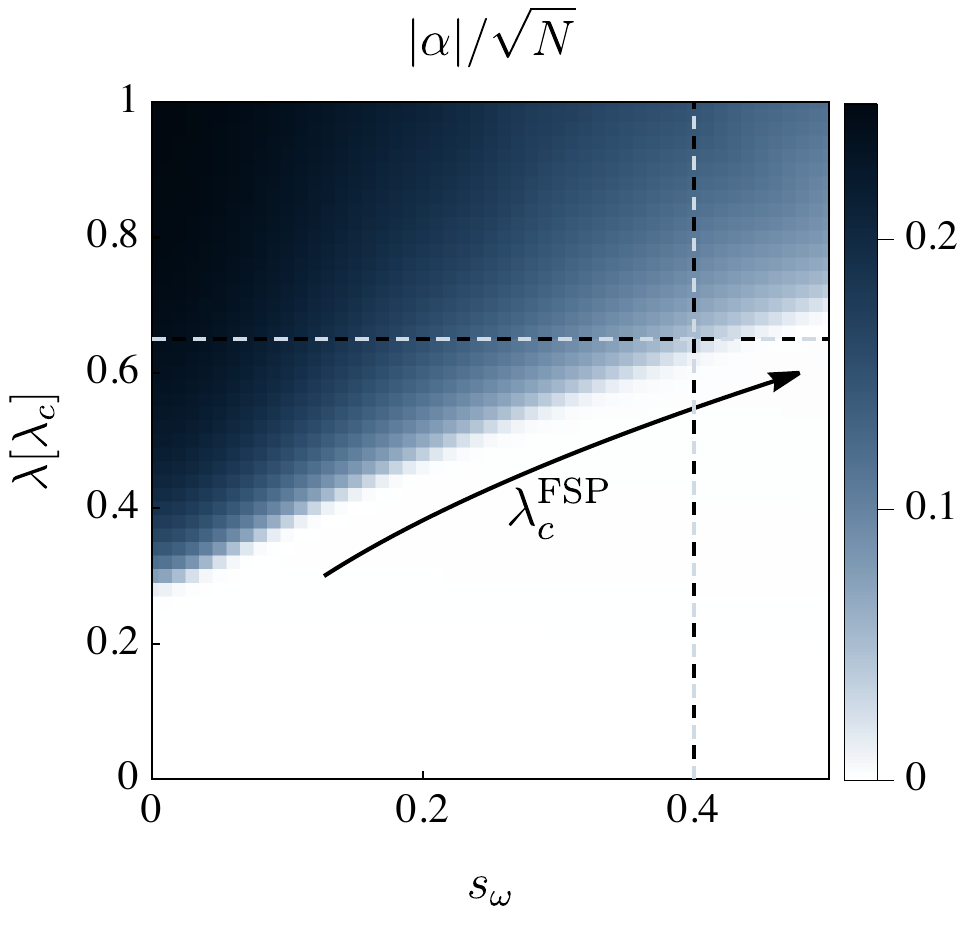}
    \caption{
    {\bf Effect of inhomogeneous broadening on the FSP transition.}
    The FSP transition at the onset driving field strength as a function of the coupling strength $\lambda$ and the inhomogeneous broadening parameter $\sigib$.
    The vertical dashed line corresponds to the case that we show in Fig.~\ref{floquet}.
    }
    \label{detuning}
\end{figure}

In Fig.~\ref{detuning}, we show the amplitude of the light field across the FSP transition as a function of the inhomogeneous broadening standard deviation $\sigib$.
We find that the transition remains sharp for increasing $\sigib$.
However, the value of $\lambda$ at which the FSP transition occurs increases with $\sigib$ while the amplitude $|\alpha|$ of the light field decreases.
This is a consequence of a subset of TLSs not contributing to the FSP due to being far detuned from the cavity frequency. 
To compensate for this lack of contributing TLSs, the coupling strength needs to increase in order to enter the FSP.
If the detuning due to inhomogeneous broadening is larger than the intrinsic linewidth of a given TLS, it is not affected by the effective driving that is present due to the interaction with the cavity which contains a non-zero light field.

In order to gain insight into the Floquet states of the system as well as their occupation, we consider the TLSs embedded in a larger space spanned by the creation (annihilation) operators $b_{1,2}^{j,(\dagger)}$.
They are connected to the Pauli matrices via
\begin{align}
    \sigma^j_x&= b_2^{j,\dagger} b^j_1 + b_1^{j,\dagger} b^j_2\\
    \sigma^j_y&=i(b_2^{j,\dagger} b^j_1 - b_1^{j,\dagger} b^j_2)\\
    \sigma^j_z&= b_2^{j,\dagger} b^j_2-b_1^{j,\dagger} b^j_1.
\end{align}
We note that the operators $b_{1,2}$ can be either fermionic operators or hardcore bosons. 
While we introduce them as auxiliary operators in this work, a natural platform to implement the dynamical state that we put forth here, is in a two-band material. 
In that implementation, the electron operators of the two bands coincide with the Schwinger operators that we use here.
In this Schwinger representation, individual two-time correlation functions $\braket{b^{j,\dagger}_n(t_2)b^{j}_n(t_1)}$ are accessible, rather than propagators of particle conserving operators Eq.~14--16. 
We can calculate the steady state distribution
\footnote{Eq.~17 can be transformed such that it is always $t_2>t_1$. Then the correlation function in Eq.~18 can be calculated by strict forward propagation using the master equation and applying the operators $b_n^j$ and $b_n^{j,\dagger}$ at $t_1$ and $t_2$, respectively.}
\begin{equation}
    n_j(\omega) = \frac{1}{(t_\mathrm{b}-t_\mathrm{a})^2}\int_{t_\mathrm{a}}^{t_\mathrm{b}} \int_{t_\mathrm{a}}^{t_\mathrm{b}} \mathcal{G}^j(t_2,t_1) e^{i\omega(t_2-t_1)}dt_2dt_1
\end{equation}
for each of the $N$ TLSs with the correlation function
\begin{equation}
    \mathcal{G}^j(t_2,t_1) = \sum_{n=1}^2\braket{b_n^{j,\dagger}(t_2)b^j_n(t_1)}.
\end{equation}
Here $t_\mathrm{a}$ is a time that is large enough for the system to have formed a steady state and 
$t_\mathrm{b}-t_\mathrm{a}$ is an interval that is large enough to ensure sufficient frequency resolution.
The distribution $n_j(\omega)$ reveals the energies and occupation of the Floquet states of the $j$th TLS.
Collecting the steady state distributions of all $N$ TLSs gives the collective distribution of the entirety of TLSs
\begin{equation}
    n(\omega) = N^{-1}\sum_{j=1}^N n_j(\omega),
\end{equation}
which displays the frequency resolved distribution of the ensemble of TLSs, rather than that of individual TLSs.
Taking the difference of the collective distribution across opposite frequencies gives the relative collective distribution 
\begin{equation}
    \Delta n(\omega) = n(\omega)-n(-\omega).
\end{equation}
$\Delta n(\omega)$ displays the frequency resolved imbalance of the ensemble and reveals the effective population inversion of the entire system and how it is depleted in the FSP. 

\begin{figure}
    \centering 
    \includegraphics[width=1.0\linewidth]{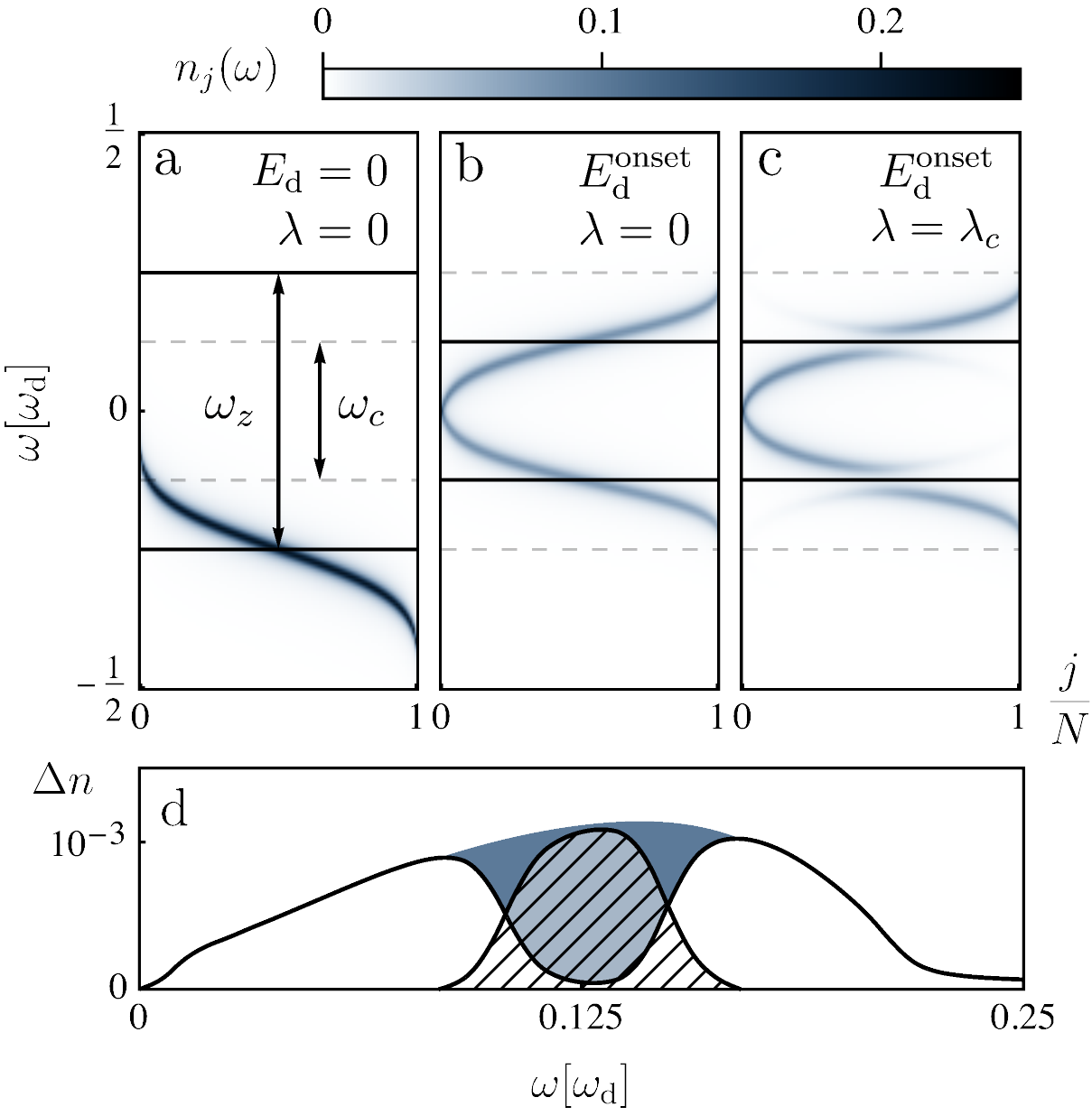}
    \caption{
        {\bf Two-level steady state distributions in the presence of inhomogeneous broadening. }
        Panels (a) through (c) show the steady state distributions $n_j(\omega)$ for $N=100$ TLSs at $\ed=0$ (a) and $\ed=\ed^\mathrm{onset}$ (b, c) as well as $\lambda=0$ (a, b) and $\lambda=\lambda_c$ (c).
        The distributions are concentrated at the Floquet energies of the broadened energy levels. 
        The horizontal lines show the average level spacing $\omega_z$ and the cavity frequency $\omega_c$.
        Panel (d) shows the relative collective distribution $\Delta n(\omega)$ for $\lambda=\lambda_c$ (black line, white filling) and $\lambda=0$ (dark blue filling) the difference between the two (hatched filling) is the effective population inversion of Floquet states that is depleted to sustain the FSP. 
    }
    \label{floquet}
\end{figure}

In Figs.~\ref{floquet}~(a) through (c), we show the steady state distributions $n_j(\omega)$ for $\sigib=0.4$ and different combinations of $\ed$ and $\lambda$.
The horizontal lines indicate the centered two-level spacing $\omega_z$ and the cavity frequency $\omega_c$.
The solid line indicates the centered spacing of the Floquet levels that for the $j$th TLS have the energy 
\begin{equation}
    \epsilon_F^j = \pm\left(\frac{\omd}{2} - \sqrt{\frac{\ed^2}{\omd^2} + \frac{(\omd-\omega_z^j)^2}{4}}\right).
    \label{floqene2}
\end{equation} 
In Fig.~\ref{floquet}~(a), we show the case of $\ed=0$ and $\lambda=0$.
In this equilibrium state the lower levels are all fully populated with energies given by Eq.~\ref{erf}.
In Fig.~\ref{floquet}~(b), we show the case of $\ed=\ed^\mathrm{onset}$ and $\lambda=0$.
As expected, the states are dominantly distributed around the Floquet energies given by Eq.~\ref{floqene2}.
At this driving field strength the Floquet states are population inverted.
In Fig.~\ref{floquet}~(c), we show the case $\ed=\ed^\mathrm{onset}$ and $\lambda=\lambda_c$.
This case is inside the FSP and the Floquet states that are close to resonant with the cavity frequency are modified due to the presence of the finite photon field in the cavity.
Hence, a gap opens and the populations of the Floquet states are modified.

In Fig.~\ref{floquet}~(d), we show the relative collective distribution $\Delta n(\omega)$ for $\sigib=0.4$, $\ed=\ed^\mathrm{onset}$ and for both $\lambda=0$ (dark blue filling) and $\lambda=\lambda_c$ (black line, white filling), which correspond to Figs.~\ref{floquet} (b) and (c), respectively.
We see that inside the FSP and close to the cavity resonance, the Floquet states are not only modified by a gap opening, but the effective population inversion is largely depleted.
The inhomogeneous broadening leads to off-resonant TLSs that are unaffected and do not contribute to the FSP mechanism.
This type of hole-burning is responsible for the reduced light field amplitude as a function of $\sigib$ in Fig.~\ref{detuning}.

\section{Two-Level and Cavity Losses}

\begin{figure}
    \centering 
    \includegraphics[width=1.0\linewidth]{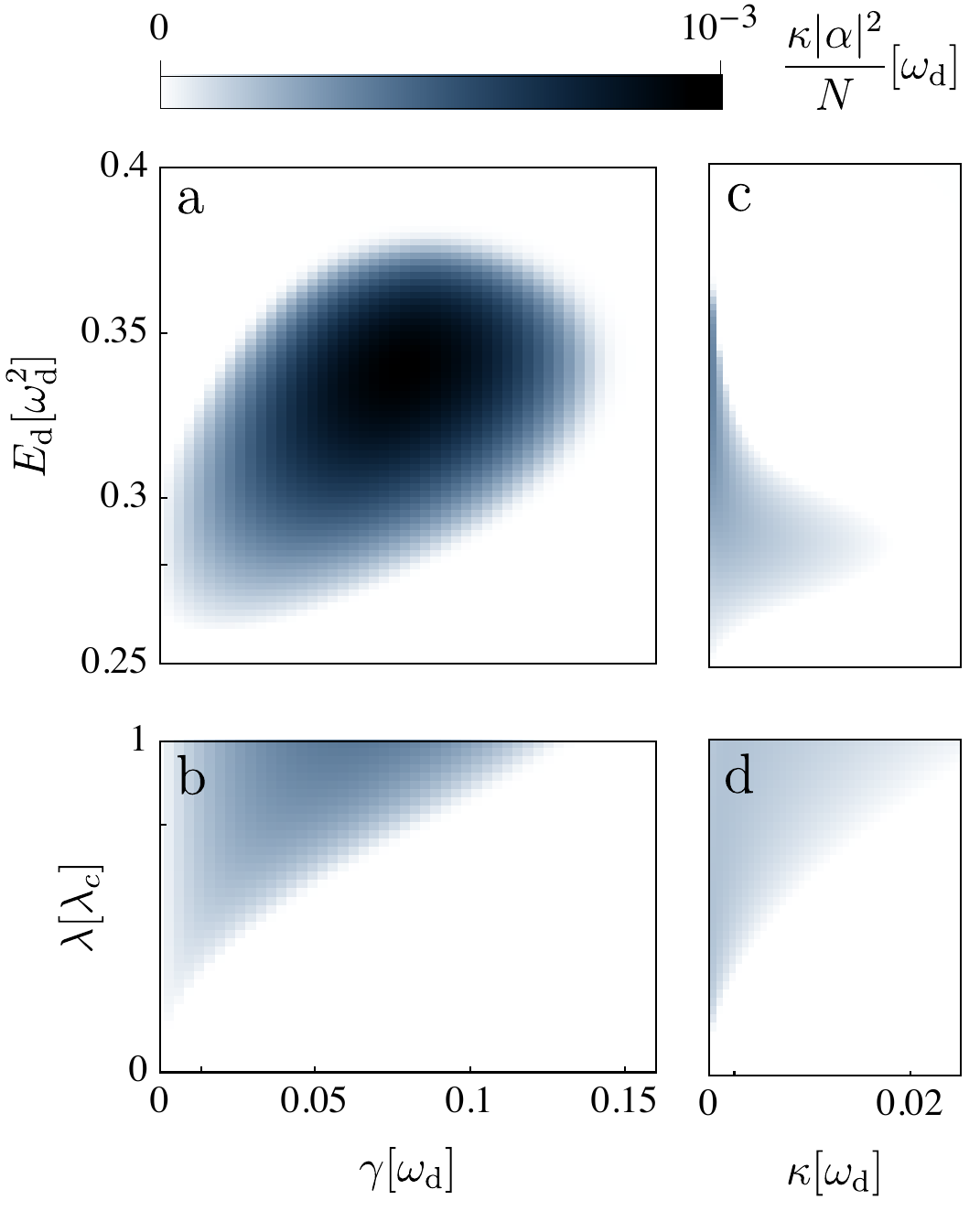}
    \caption{
        {\bf Effects of strong dissipation on the FSP.}
        The output intensity of the light field per TLS $\kappa |\alpha|^2/N$ as functions of combinations of $\ed$, $\lambda$, $\kappa$ and $\gamma$.
        In panel (a), we show the output intensity per TLS as a function of $\gamma$ and $\ed$ for $\lambda=0.75\lambda_c$ and $\kappa=\omd/400$.
        In panel (b), we show the output intensity per TLS as a function of $\gamma$ and $\lambda$ for $\ed=\ed^\mathrm{onset}$ and $\kappa=\omd/400$.
        In panel (c), we show the output intensity per TLS as a function of $\kappa$ and $\ed$ for $\lambda=0.75\lambda_c$ and $\gamma=\omd/24\pi$.
        In panel (d), we show the output intensity per TLS as a function of $\kappa$ and $\lambda$ for $\ed=\ed^\mathrm{onset}$ and $\gamma=\omd/24\pi$.
    }
    \label{dissip}
\end{figure}

As a third metric for robustness, we show the effect of the TLS dissipation rates and the cavity loss rate on the FSP.
In Fig.~\ref{dissip} we show the magnitude of the light field in the cavity as a function of the effective driving field strength $\ed$ and the coupling strength $\lambda$, as well as the dissipation coefficients.
The quantity we show is the intensity per TLS of the light field $|\alpha|^2 N^{-1}$ multiplied with the cavity loss rate $\kappa$. 
This serves the purpose of quantifying the coherent output of the system rather than the photon count inside the cavity.

In Fig.~\ref{dissip}~(a) and (b) we show the effect of the TLS dissipation coefficients.
We choose the example $\gamma=\gamma_z=2(\gamma_-+\gamma_+)$, i.e. we keep the ratio of $\gamma_-+\gamma_+$ and $\gamma_z$ constant for convenience.
We see that the FSP transition is robust for increasing values up to $\gamma\lesssim 0.15 \omd$ for $\lambda=0.75\lambda_c$.
In our example, that is inspired by light-driven graphene, this corresponds to $\gamma\approx 45\si{\tera\hertz}$ compared to the characteristic energies of the Hamiltonian of the order of $\omd=2\pi\times48\si{\tera\hertz}$.
This demonstrates a robustness to dissipation up to rates which are comparable to the coefficients used to describe experimental setups of driven graphene \cite{Elsaesser11,Cavalleri13,Polini13,Kurz14,McIver,Hommelhoff21}.
This further supports the viability of graphene as a potential platform for hosting the FSP.

In Fig.~\ref{dissip}~(c) and (d) we show the effect of the cavity loss rate $\kappa$ on the FSP. 
We find a small range of available values for $\kappa$ at which the FSP emerges and further the threshold at which the FSP becomes unsustainable is comparatively small at values of $\kappa \lesssim 0.02\omd$ for $\lambda=0.75\lambda_c$.
In our example this corresponds to $\kappa\approx 5\si{\tera\hertz}$.
This demonstrates the sensitivity of the FSP with respect to cavity losses, which we identify as the most sensitive parameter. 

For the purpose of realizing the FSP, for driving field strengths close to $\ed^\mathrm{onset}$ the FSP will occur for large values of $\kappa$, if $\lambda$ is sufficiently large. 
Similarly, the given magnitude of $\lambda$ provides an upper limit on $\kappa$ to achieve the realization of the FSP. 
We note that the FSP can be realized for arbitrarily small values of $\lambda$, given sufficiently small $\kappa$. 
Therefore, for a given platform, a high-finesse cavity might enable the realization. 
We note, however, that the optimal output intensity is achieved for intermediate magnitudes of both $\kappa$ and $\lambda$. 
Stronger dissipation results in a larger inversion of the TLSs, and therefore in a higher output intensity. 
So for the optimal operation of FSP, intermediate values of the dissipation rates are desirable.

\section{Conclusion}

We have demonstrated the robustness of the Floquet-assisted superradiant phase (FSP) in the parametrically driven Dicke model with a dissipative model, that is designed for the electron dynamics in a solid.
The photonic steady state in the FSP is robust against inhomogeneous broadening, reasonably strong dissipative processes of the two-level systems and phase diffusion in the driving field.
Across the FSP transition, the linewidth of the light field narrows drastically and overcomes the linewidth of the driving field as well as the intrinsic linewidth of the cavity that is given by its loss rate, which is a hallmark of laser mechanisms.

In our model the dissipation is performed in the instantaneous eigenbasis.
This choice is motivated by the capabilities of capturing dynamics of two-band solids, as has been demonstrated in previous work \cite{Nuske}.
The dependence of the FSP on the ratio of characteristic frequencies and temperature, as well as the driving field strength in units of the driving frequency squared suggests the most promising range of driving frequencies to be of the order of tens to hundreds of terahertz.
This energy scale also coincides with the situation in two-band solids such as light-driven graphene \cite{McIver,Nuske,Broers}.
The values of the dissipation coefficients up to which we find the FSP to be stable also agrees with realistic estimates of coherence times in available graphene samples \cite{Elsaesser11,Cavalleri13,Polini13,Kurz14,McIver,Hommelhoff21}. 

We conclude from the robustness and the small values of the critical coupling of the FSP that it can be utilized for laser operation. 
The FSP is in principle accessible in two-band solids such as light-driven graphene under realistic conditions.
The details of the collective effect of solid-state dispersion relations on the emergence of the FSP will be the subject of future research.
Such a demonstration would then constitute a Floquet-assisted solid-state laser system in the terahertz frequency domain. 
As such, we propose to expand the range of dressed-state laser mechanisms into the domain of Floquet-engineered electron bands in solids, that is accessible with current pump-probe technology. 

\section*{Acknowledgements} 
This work is funded by the Deutsche Forschungsgemeinschaft (DFG, German Research Foundation) -- SFB-925 -- project 170620586,
and the Cluster of Excellence 'Advanced Imaging of Matter' (EXC 2056), Project No. 390715994.

\bibliography{lit}
\end{document}